# Detecting and modelling real percolation and phase transitions of information on social media

Jiarong Xie[1], Fanhui Meng[1], Jiachen Sun[1], Xiao Ma[1], Gang Yan[2,3], Yanqing Hu[1,4*]

[1]School of Computer Science and Engineering, Sun Yat-sen University, Guangzhou 510006, China.
[2]School of Physics Science and Engineering, Tongji University, Shanghai 200092, China.
[3]Shanghai Institute of Intelligent Science and Technology, Tongji University, Shanghai 200092, China.
[4]Southern Marine Science and Engineering Guangdong Laboratory, Zhuhai 519082, China.

* Corresponding author: Yanqing Hu (huyanq@mail.sysu.edu.cn; https://orcid.org/0000-0003-0653-0259)

**Abstract**

**It is widely believed that information spread on social media is a percolation process, with parallels to phase transitions in theoretical physics. However, evidence for this hypothesis is limited, as phase transitions have not been directly observed in any social media. Here, through analysis of 100 million Weibo and 40 million Twitter users, we identify percolation-like spread, and find that it happens more readily than current theoretical models would predict. The lower percolation threshold can be explained by the existence of positive feedback in the coevolution between network structure and user activity level, such that more active users gain more followers. Moreover, this coevolution induces an extreme imbalance in users' influence. Our findings indicate that the ability of information to spread across social networks is higher than expected, with implications for many information spread problems.**

**Introduction**

Online social media create large-scale networks that enable information spreading to occur rapidly and widely [1-11] and play an important role in a variety of applications [1,10,12-31]. The spread of information on a social network is a cascading process. A node (user) generates an information item, and its followers retweet the item to their own followers with a probability. The process continues until no new node receives the item, forming an information cascade. In this emerging interdisciplinary field, a widely used hypothesis [2,8,32,33] is that information cascading is a dynamical process of percolation, connecting the researches of popular social media to the fundamental theories of phase transition in theoretical physics. This triggers a series of impactful theoretical results on viral marketing [18,34], optimal percolation [2,8,32,33] and rumor containment [35-37]. Despite the significant advances, this hypothesis has not been verified or falsified, and the phase transition predicted by previous theories has not been directly observed in any real large-scale social media. The major challenge lies in the fact that, in order to calculate the order parameter of percolation phase transition, one must analyze a large number of information spreading trajectories, each of which requires not only the friendship data of the users who have forwarded but also of other users who received the information item. Due to small-world-ness [38] and scale-free-ness [39] of social media networks, the information tracks with large cascading size contain several hub nodes who can have millions of followers. Hence, to avoid bias, it is essential to collect almost the whole dynamic network data in a short period of time.

**Results**

Here we address the challenging problems in information spreading, by empirically assessing whether information spreading on online social networks can be effectively modeled by site percolation (see Supplementary Figures 1-3 and Supplementary Discussion 1 for discussions on other percolation models). Building on that, we quantitatively determine the threshold of information outbreaks with huge datasets that contain social networks of approximately 100 million (41 million) nodes and the activities of up to 0.3 million nodes in Weibo (Twitter) (see Tables. 1 and 2, Supplementary Figure 4 and Supplementary Result 1 for details). Site percolation model was originally used to study the size, distribution and phase transition of the connected components after certain nodes are removed. More recently, it has been widely used in the research of information dissemination or infectious diseases propagation. By adopting this framework, we assume that for each message, certain nodes are interested in the message and will retweet after they receive it (considered as occupied nodes), while other nodes will not (considered as removed nodes). If the retweeting (occupation) probabilities are the same for all nodes, we call it uniform site percolation (see Fig. 1A), otherwise we call it heterogeneous site percolation (see Fig. 1B). In this framework, the collective group of retweeting nodes in a global information cascade corresponds to the giant out-component (GOUT) in site percolation [40,41]. See Supplementary Figures 5-7, Supplementary Table 1 and Supplementary Discussion 2 for detailed discussion of the relation between site percolation and information spreading.

For a given message in a social network with size $N$, we define $\beta$ as the fraction of occupied nodes in the network, which is essentially the retweet probability, i.e., if all users in social network receive the information item, $\beta$ of them will retweet. We use $P_\infty(\beta)$ to denote the fraction of nodes who retweet the message in the whole network of $N$ nodes, and $P_e(\beta)$ to denote the fraction of exposed nodes who receive the message. If the information item is retweeted numerous times, we have large $P_\infty$ and $P_e$ values, and it is reasonable to assume that $P_\infty$ is a representative sample of all willing-to-retweet users and $P_e$ is a representative sample of all users. Then $P_\infty/P_e$ can be used to estimate the value of $\beta$ (see Extended Data Fig. 1, Supplementary Figure 8, Supplementary Method 1 and Supplementary Discussion 3 for details). As shown in Fig. 1C, we find that 98.4% of messages that are empirically observed to be in global outbreak are wrongly predicted to be only having limited local spreadings according to the previous uniform site percolation model [41,42] (see Supplementary Discussion 3 for discussion of cascade data, Extended Data Fig. 2 and Supplementary Discussion 4 for cascade states). Moreover, the threshold for information outbreaks is only about one tenth of the predicted value given by the uniform site percolation (Fig. 1D). These findings suggest that previous theoretical studies assuming uniform spreading probability $\beta$ might significantly underestimate the outreach of information in social media networks (see Extended Data Fig. 3, Supplementary Figures 9-12, Supplementary Table 2 and Supplementary Discussions 5 and 6 for detailed discussions of previous models).

The significant difference between the actual and the previously predicted threshold prompts us to quantitatively explore the interplay between information tracks and network structure. We characterize each node by the number $k_o$ of its followers, the number $k_i$ of its followees, and its activity $m$ defined as $m = \frac{\Delta s}{\Delta t}$, where $s$ is the total number of its tweets. Hence, $m$ is the average number of tweets that the node posts within a unit of time $\Delta t$. Our systematical analysis of $k_o$, $k_i$ and $m$ of each node reveal positive correlations between $m$ and $k_o$, a phenomenon qualitatively consistent with previous results [43-46]. This is indirectly related to Ref. [45], in which Antoniades and Dovrolis track the diffusion trajectory of tweets and find that the receiver of the tweet is more likely to become a new follower of the original poster of the tweet. Moreover, we perform quantitative analyses on the data in three different periods, years 2012-2014, and find that the correlation is time-varying and can be characterized by a power function,

$$m \sim k_o{}^\alpha. \tag{1}$$

Within each period, the power function has different exponents in two regimes (Figs. 2A and 3A), i.e., the activity of the nodes who have a great many followers tend to increase slowly or even decrease in Twitter and Weibo. More importantly, we continu-ously collect the data of network evolution during the three years and discover a positive-feedback effect in the coevolution between network structure and nodal activity,

$$\frac{\Delta k_o}{\Delta t} \sim k_o^{\sigma} m^{\tau} \qquad (2)$$

(see Figs. 2B and 3B, Supplementary Figure 13 and Supplementary Result 2 for details). The positive feedback effect makes the correlation Eq. 1 become stronger with time. Indeed, as shown in Fig. 2A, the value of $\alpha$ for Weibo increases from 0.158 ($\pm 0.019$) in year 2012 to 0.284 ($\pm 0.017$) in year 2014 (see Tab. 2 for all the values of $\alpha$ within different periods in Weibo and Twitter). The sampling process, statistical methods and biases are discussed in Supplementary Figures 14-19, Supplementary Tables 3-10, Supplementary Result 1 and 3, and Supplementary Method 2. We will show later that the increasing of $\alpha$ leads to a lower threshold of information cascades and more serious imbalance in the influence of users on social media networks. In contrast, the correlation between $m$ and $k_i$ is negligible (see Supplementary Figures 20 and 21, and Supplementary Result 4). Since the distributions of $k_o$ in both networks are heavy-tailed (see Supplementary Figure 4 and Supplementary Result 1), Eq. 1 means that nodal activities are heterogeneous, i.e., the nodes with more followers post more frequently; Eq. 2 indicates that the growth rate of followers is also highly variable, i.e., the number of a node's followers increases faster if this node has more followers or posts more frequently.

The positive-feedback coevolution mechanism plays an important role in dynamical processes taking place on social media networks. Here we focus on its impact on information cascades. An action of posting or retweeting an information item could attract new followers who might also retweet the item, yet the timescale of such structure evolution is much longer than the duration of an information cascade. Hence it is reasonable to assume that the structure of the social network is kept unchanged when a specific information item propagates over it. However, different from the widely used assumption [40,47-49], Eq. 1 implies that the retweeting probabilities of nodes are non-uniform. We incorporate nodal activity into network site percolation theory and perform data-driven simulations of information cascades. In the simulations, a retweet probability $\beta$ represents the proportion of nodes on the network who will retweet if receiving the information, characterizing the intrinsic popularity of an information item. We pick $\beta N$ nodes on the Weibo network, not randomly but with probabilities $q$ proportional to nodal activity (see Supplementary Method 3),

$$q \propto m(k_i, k_o). \qquad (3)$$

The size of the giant-out component formed by these $\beta N$ nodes corresponds to the size of an information cascade [41,42,49].

It is worth noting that Eq. 3 is a general form describing the dependence of nodal activity on the numbers of followers and followees. Yet, the correlation between $m$ and $k_i$ does not significantly alter the results of data-driven simulations that considers only Eq. 1 (see Table 3 and Supplementary Figure 21). In contrast, ignoring the correlation between $m$ and $k_o$ leads to predictions close to the outcome of the uniform percolation model [41,42], significantly deviated from empirical data (see Supplementary Figure 21). Hence, we use correlation Eq. 1 to investigate the impact of $\alpha$ on cascading threshold. As shown in Figs. 2C and 3C, when $\alpha$ increases with time due to the positive-feedback effect, the cascading threshold decreases and hence the capacity of social media to spread information becomes higher. Indeed, the threshold of empirical cascades collected in October 2013 agrees better with our percolation model than the percolation with uniform retweeting probability ($\alpha = 0$) (see Figs. 2G-I and Extended Data Fig. 4,

Supplementary Figures 22-24, Supplementary Table 11 and Supplementary Result 5). The agreement between the empirical data and models, sampling bias and correlation coefficient measurement methods are discussed in Fig. 2D, Supplementary Figures 25-27, Supplementary Method 2, Supplementary Results 5 and 6. Although information cascades are widely believed as complex contagion [50,51] and depend on many microscopic (often unknown) details, our findings suggest that network structure and the relationship Eq. 1 are two main factors that determine the threshold of information cascades and the spreading capacity of social media.

Next we analytically formulate the impact of the positive-feedback coevolution between network structure and nodal activity on the cascade threshold. The degree distribution of the network is denoted by $P(k_i, k_o)$. The proportion $\beta$ of nodes are picked according to Eq. 3. Then the degree distribution of the picked nodes becomes

$$\tilde{P}(k_i, k_o; \beta) = \frac{1}{\beta} P(k_i, k_o)[1 - t^{m(k_i,k_o)}], \tag{4}$$

in which $t$ is the solution of equation (see Methods section)

$$\sum_{k_i,k_o} P(k_i, k_o)[1 - t^{m(k_i,k_o)}] = \beta. \tag{5}$$

Defining the generation function

$$G(x, y) = \sum_{k_i,k_o} \tilde{P}(k_i, k_o; \beta) x^{k_i} y^{k_o}, \tag{6}$$

then the giant out-component size $P_\infty$, which is also the cascade size, can be obtained through

$$P_\infty = \beta - \beta G(1 - \theta, 1). \tag{7}$$

The $\theta$ satisfies the self-consistent equation $\theta = \frac{\beta[G_y(1,1) - G_y(1-\theta,1)]}{\langle k \rangle}$, where $G_y(x, y) = \frac{\partial G(x,y)}{\partial y}$. The cascade threshold $\beta_c$ is

$$\beta_c = \sum_{k_i,k_o} P(k_i, k_o)[1 - t_c^{m(k_i,k_o)}], \tag{8}$$

where $t_c$ is the solution of equation $\sum_{k_i,k_o} k_i k_o P(k_i, k_o) [1 - t_c^{m(k_i,k_o)}] = \langle k \rangle$ (see Methods section and upplementary Figures 28 and 29, and Supplementary Method 4 for the detailed derivation).

We validate the theoretical analyses by comparing its results with that obtained by data-driven simulations. As shown in Figs. 2C and 3C, our theory is in good agreement with the simulations on real social media and synthetic networks (see Supplementary Figure 29). In particular, as the exponent $\alpha$ in Eq. 1 increases, i.e., the heterogeneity of retweeting probability becomes higher, the cascade threshold $\beta_c$ decays approximately exponentially, leading to a dramatic reduction of the cascading threshold in both Weibo and Twitter (Fig. 3D). When $\alpha = 0$, this model degenerates to the classical percolation with uniform retweeting probability. Moreover, the heterogeneity of retweeting probability in our model does not change the critical behavior of percolation, i.e., the critical exponent of order parameter near the percolation threshold equals 1, agreeing well with empirical data (Fig. 2E, and see also Supplementary Figure 30, Supplementary Methods 4 and 5, and Supplementary Discussion 7).

Social media are open platforms that enable everybody to express his/her opinions, with a potential of shifting mass communications dominated by centralized media organizations to interpersonal communications between individuals. The above quantitative analyses show that users' retweeting probabilities tend to be more heterogeneous, which leads to lower thresholds of information cascades. However, there might be inequality in the abilities of users to trigger information cascades. Such ability indicates the influence of users in social media and has the profound impact in shaping public opinions on political, economic and societal issues [52-54]. To investigate the evolution of users' influence distribution we randomly sample 11 thousands of users from a set of 100 million users in Weibo and collect all of their information tracks from their registration dates (see Supplementary Figure 5) to the end of 2014, which contains 9.4 million posts in total (see Supplementary Figures 31 and 32, and Supplementary Discussion 8). The influence of user $j$ is calculated as

$$w_j = \frac{M_j}{M},\qquad(9)$$

where $M_j$ denotes the number of outbreak information items in a period of time that are originally posted by user $j$, and $M = \sum_j M_j$ denotes the number of all outbreak information items (cascading size > 2000, see Supplementary Figure 31) in the same period of time. We empirically and theoretically find that the users influence becomes more concentrated as $\alpha$ increases with time (Figs. 2F and 3E). In fact, in year 2014 the top 0.7% of users possess 99.3% of influence power (see Supplementary Result 1C and Supplementary Discussion 8 for more information about relevant dataset, methodology and bias analysis). In contrast, the uniform theory predicts that top 12% of users possess 88% of influence power. While elite users inevitably have more influence power [52,53], the imbalance in real social networks is 17 times more serious than the prediction of uniform percolation. This finding implies that current social media, built with the intention of enabling people to freely express themselves, are actually not as decentralized as expected. The voice of an extremely large proportion of common users is still suppressed. Even worse, our empirical and analytical results indicate that, due to the positive-feedback effect, the imbalance in influence of users tends to become more serious as social media systems evolve.

**Discussion**
In conclusion, we empirically confirm that information spreading can be effectively modeled by heterogeneous site percolation and detect the corresponding second-order phase transition on social media networks. Moreover, we reveal the positive-feedback mechanism in the coevolution between network structure and user activity, and it leads to an unexpectedly high capacity of information spreading and extreme imbalance in influence power of users. Importantly, our findings indicate that it is necessary to rethink a couple of problems pertaining to information cascades, such as influence maximization, social contagion and protocol design of social media that were usually analyzed on static networks. It is important to note that while our model captures the main factors that determine the dynamical properties of information cascades, there are some variations in the trend across different years and different datasets, especially in the correlation between network structure and nodes' activity level over time. We find it is difficult to have a simple mechanism that can explain such variations, as it could be due to some hidden factors not observed from the data itself. In general, the empirical evidence suggests that this mechanism is similar to the preferential attachment but is more complex. It is worth exploring in future studies.

**Methods**
**Datasets.** The Weibo data contains three datasets: (i) almost the whole friendship network of 100 million users, (ii) the evolution data of user profile, which contains a three-year evolution of follower count, followee count and tweet count of at least 185 thousand users, and (iii) tracks of 253 real outbreak information. The Twitter data contains a network [43] with about 41 million users and the evolution data of about 184 thousand users. The data (both Weibo and Twitter) were collected and used in compliance with the terms for those platforms. The network and evolution datasets are summarized in Tab. 1 and 2 respectively. More details of the datasets are shown in Supplementary Result 1. No statistical methods are used to pre-determine sample size but our samples are the largest dynamic datasets that we have tried our best to obtain. The binning method is employed to quantify the nodal activity and follower growth rate. If the number of users in a bin is small (the cutoff is 10 in Weibo), the bin is excluded from the analysis to avoid large fluctuation.

**Description of percolation models.** The data-driven percolation models are summarized in Tab. 3. By using the nodal activity data in different years, we can model the information spreading in the corresponding year. For model M1 and M6, the nodal activity takes its actual value if its activity data is recorded in our datasets, otherwise it takes the activity value of a randomly picked node whose follower count is closest.

Using roulette wheel selection method [55], the $\beta N$ nodes are picked one by one without repetition and at each time the probability $q_j$ of picking node $j$ is proportional to its activity: $q_j = \frac{m_j}{\sum_l m_l}$, where the $m_j$ denotes the activity of node $j$ and the denominator is the sum of the nodes that have not been picked. The $\beta N$ picked nodes constitute a sub-network $\mathcal{G}'$ of the original network $\mathcal{G}$, and the edges connecting the $\beta N$ picked nodes on $\mathcal{G}$ are preserved on $\mathcal{G}'$. We then identify the giant connected components of this sub-network $\mathcal{G}'$. Since the nodes of the GOUT indicate the retweeting users of the information item, therefore the GOUT size is equivalent to the cascade size $P_\infty$.

**Theoretical solution of data-driven percolation model.** In order to get the cascade size and cascade threshold, we first study the picked nodes. We use $\widetilde{N}(k_i, k_o; \beta)$ to denote the number of the remaining nodes (i.e., the nodes that have not been picked) with degree $(k_i, k_o)$ after a proportion $\beta$ of nodes are picked up. Notice that, the degree $(k_i, k_o)$ of nodes here is the degree on the original network. Here, we employ the target attack percolation [55] to solve our data-driven cascading model. When a node is newly picked, $\beta$ increases by $1/N$ and $\widetilde{N}(k_i, k_o; \beta)$ changes as

$$\widetilde{N}\left(k_i, k_o; \beta + \frac{1}{N}\right) = \widetilde{N}(k_i, k_o; \beta) - \frac{Q(k_i, k_o; \beta) m(k_i, k_o)}{\langle m(k_i, k_o) \rangle_\beta}, \tag{10}$$

where $Q(k_i, k_o; \beta)$ denotes the degree distribution of the remaining nodes, and $\langle m(k_i, k_o) \rangle_\beta = \sum_{k_i, k_o} Q(k_i, k_o; \beta) m(k_i, k_o)$ is the average activity of the remaining nodes. When $N$ is large, Eq. 10 can be recast as

$$\frac{1}{N} \frac{d\widetilde{N}(k_i, k_o; \beta)}{d\beta} = -\frac{Q(k_i, k_o; \beta) m(k_i, k_o)}{\langle m(k_i, k_o) \rangle_\beta}, \tag{11}$$

The definition of $\widetilde{N}(k_i, k_o; \beta)$ and $Q(k_i, k_o; \beta)$ lead to

$$\widetilde{N}(k_i, k_o; \beta) = (1 - \beta) N Q(k_i, k_o; \beta). \tag{12}$$

Substituting Eq. 12 into Eq. 11, we obtain

$$(1 - \beta) \frac{dQ(k_i, k_o; \beta)}{d\beta} = Q(k_i, k_o; \beta) - \frac{Q(k_i, k_o; \beta) m(k_i, k_o)}{\langle m(k_i, k_o) \rangle_\beta}, \tag{13}$$

With the initial condition of $Q(k_i, k_o; 0) = P(k_i, k_o)$, the solution of Eq. 13 is

$$Q(k_i, k_o; \beta) = \frac{1}{1 - \beta} P(k_i, k_o) t^{m(k_i, k_o)}, \tag{14}$$

which is equivalent to Eq. 4.

Eq. 7 is the giant out-component (corresponding to the information cascading cluster) size of sub-network $\mathcal{G}'$ constituted by the picked nodes, divided by size $N$ of original network $\mathcal{G}$. In Eq. 7, $\theta$ denote the average probability that a node $j$, which is the source node of a randomly selected edge $j \to l$ from $\mathcal{G}$, being in GOUT of $\mathcal{G}'$. The source node $j$ being in GOUT of $\mathcal{G}'$ if the following two conditions are both satisfied: (i) $j$ belongs to the $\beta N$ picked nodes, and (ii) at least one of $j$'s in-coming neighbors being in GOUT of $\mathcal{G}'$. Therefore, we obtain the following self-consistent equation using Eqs. 4-6

$$\theta = \sum_{k_i,k_o} \frac{k_o P(k_i,k_o)}{\langle k \rangle}\left[1 - t^{m(k_i,k_o)}\right]\left[1 - (1-\theta)^{k_i}\right] = \frac{\beta[G_y(1,1) - G_y(1-\theta,1)]}{\langle k \rangle}. \quad (15)$$

Then the GOUT size (i.e., the cascade size) is

$$P_\infty = \sum_{k_i,k_o} P(k_i,k_o)\left[1 - t^{m(k_i,k_o)}\right]\left[1 - (1-\theta)^{k_i}\right] = \beta - \beta G(1-\theta,1). \quad (16)$$

At the critical point (threshold), $t_c$ can be solved by following equation

$$\left.\frac{d\beta[G_y(1,1) - G_y(1-\theta,1)]}{\langle k \rangle d\theta}\right|_{\theta=0, t=t_c} = \sum_{k_i,k_o} \frac{k_i k_o P(k_i,k_o)}{\langle k \rangle}\left[1 - t_c^{m(k_i,k_o)}\right] = 1. \quad (17)$$

Substituting $t_c$ into Eq. 5, we obtain critical point $\beta_c$.

**Acknowledgements**

The authors would like to thank Ling Feng, Wenyuan Liu for the very help discussions. This work was supported by Natural Science Foundation of Guangdong for Distinguished Youth Scholar, Guangdong Provincial Department of Science and Technology (Grant No. 2020B1515020052), Guangdong High-level Personnel of Special Support Program, Young TopNotch Talents in Technological Innovation (Grant No. 2019TQ05X138) and the National Natural Science Foundation of China (Grants No. 61903385, 61773412, U1911201, U1711265 and 61971454), National Key R&D Program of China (2018AAA0101203). The funders had no role in study design, data collection and analysis, decision to publish or preparation of the manuscript.

**Author contributions**

Y.H. conceived the project. Y.H., J.X. and G.Y. designed the experiments. J.X. performed experiments and numerical modeling. J.X. and Y.H. solved the model. J.X., F.M., J.S., X.M., G.Y. and Y.H. discussed and analyzed the results. J.X., G.Y. and Y.H. wrote the manuscript.

**Competing interests**

The authors declare no competing interests.

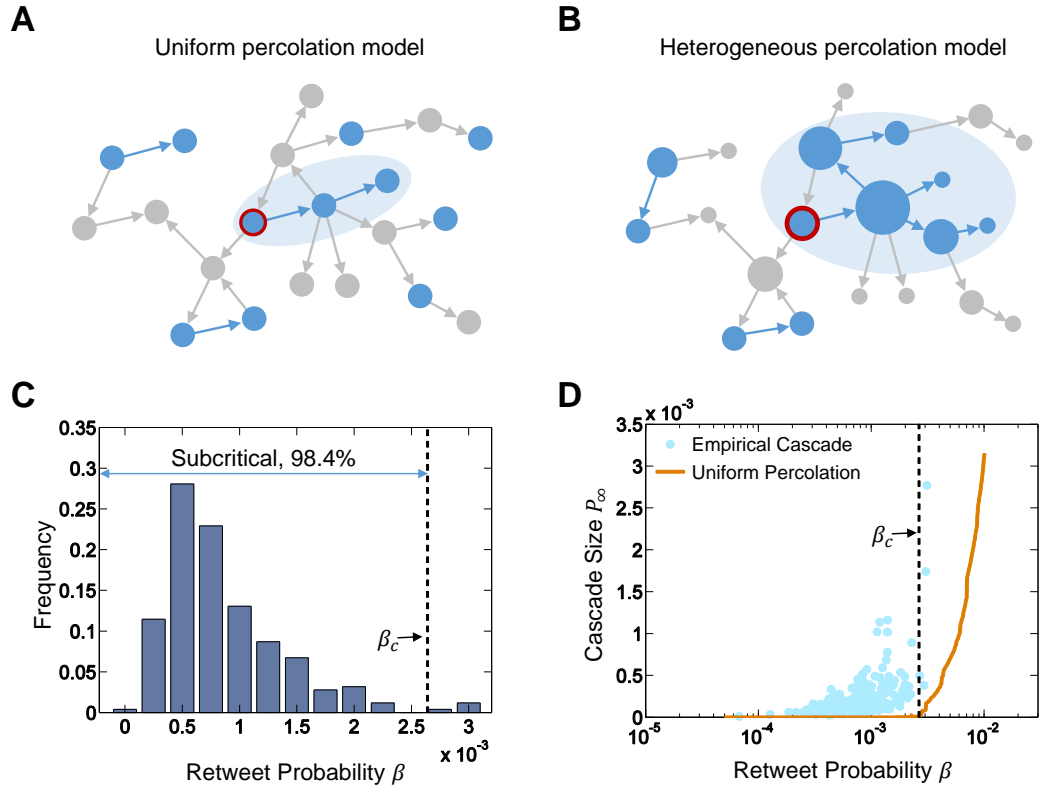

**Figure 1: Unexpectedly low threshold of information cascades in Weibo.** (**A**) Uniform percolation model. (**B**) Heterogeneous percolation model. The probability of a node being selected is positively correlated with its out degree, which is represented by the node's size. In A and B, blue nodes correspond to the occupied nodes, i.e., people will retweet the message if they receive it, and gray nodes correspond to removed nodes, i.e., people who will not retweet the message. Nodes within the shaded area are users who will retweet the message if the message is posted by the node with red frame. We can see that for a network with the same number of occupied nodes, the GOUT of heterogeneous percolation is larger. (**C**) Distribution of retweet probability $\beta$ of empirical cascades. The 98.4% of information items that actually led to cascades (see Supplementary Result 1, Discussions 3 and 5) are predicted by the uniform percolation model to be at subcritical states. (**D**) Comparison between empirical cascades and the simulation results obtained by uniform percolation model. The x-axis is plotted in log scale. Each blue dot represents a real information cascade.

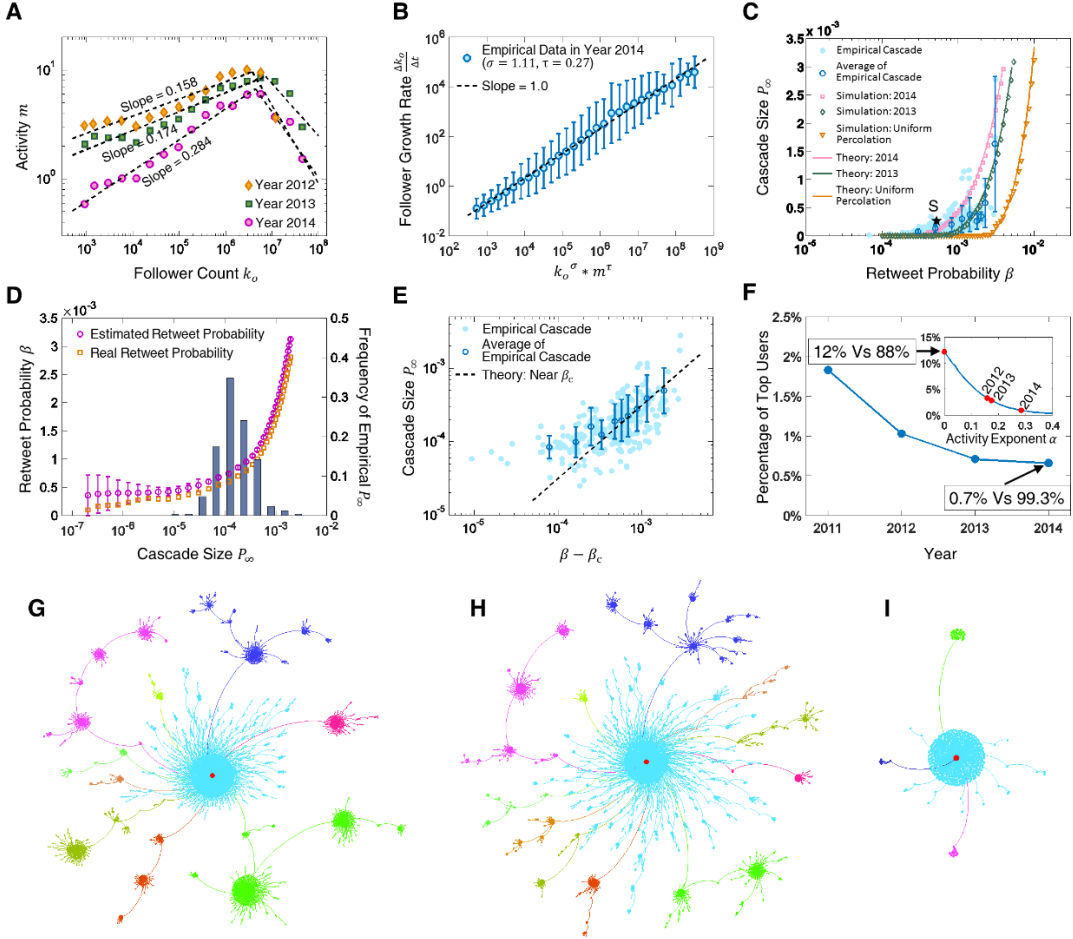

**Figure 2: Empirical results and data-driven percolation model in Weibo.** (**A**) User activity $m$ as a function of follower count $k_o$ during three different periods. Sampling bias are discussed in Supplementary Result 3. (**B**) Growth rate of follower number $\frac{\Delta k_o}{\Delta t}$ as a function of current follower count $k_o$ and activity $m$. The unit time for calculating user activity and follower growth rate is one day. (**C**) Comparison between empirical information cascades and the results obtained by data-driven percolation model (see Tab. 3). Each blue dot represents a real cascade. The pink and green symbols represent the simulation results of model M2. The curves represent theoretical results corresponding to the simulation results with the same color. (**D**) Comparison between estimated and actual retweeting probability in data-driven percolation model M2 with nodal activity in year 2014. The purple circle symbols show the relationship between cascade size $P_\infty$ and estimated retweet probability. The orange square symbols are obtained by model M2 with a given retweet probability. The histogram is the frequency of empirical $P_\infty$. (**E**) The critical behavior of model M2 in year 2014 near the cascading threshold $\beta_c$. The dashed line (slope = 1) is our theoretical result (see Supplementary Result 2B). (**F**) The percentage of top users who possess the most of influence power in Weibo. The solid curve represents the results obtained from empirical information cascades whose size is larger than 2000 (see Supplementary Discussion 8). (Inset) The percentage of top users as a function of activity exponent $\alpha$ with cascade size $P_\infty = 2000$ in model M7 in year 2014. (**G**) A typical real information track corresponds to the star S in Fig. C. The central red dot is the seed of the information, and the links represent retweeting behaviors. (**H**) and (**I**) An information track constructed by model M2 in year 2014 and uniform model M5 respectively, with the same seed and retweeting probability in Fig. G.

All error bars represent standard deviations. We use log scale for both axis in A, B and E, and set only x-axis as log scale in in C and D.

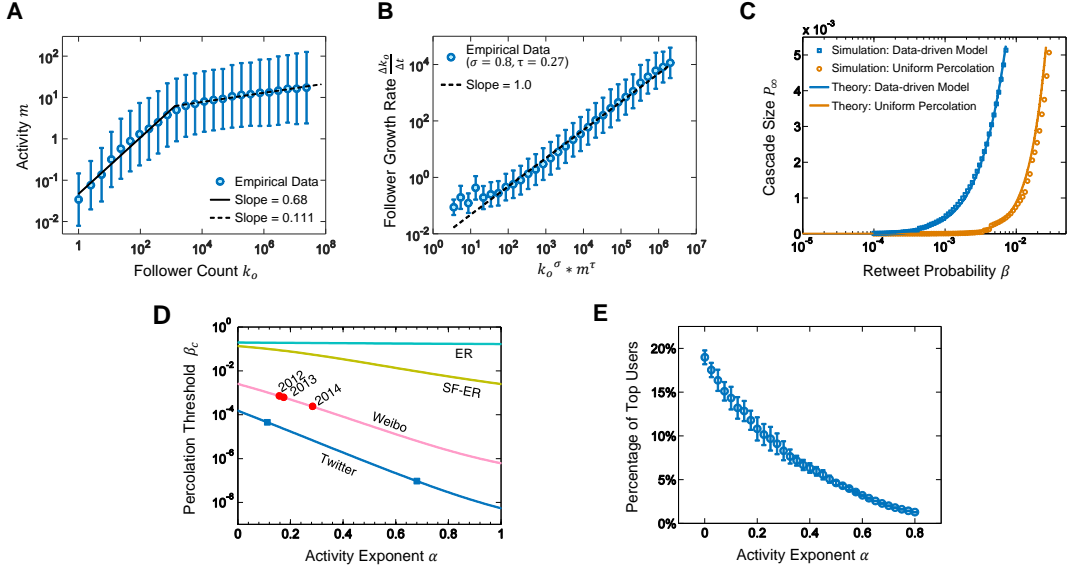

**Figure 3: Results of data-driven percolation model in Twitter.** (**A**) User activity $m$ as a function of follower count $k_o$. (**B**) Growth rate of follower number $\frac{\Delta k_o}{\Delta t}$ as a function of current follower count $k_o$ and activity $m$. The unit time for calculating user activity and the growth rate of the follower number is one day. (**C**) Comparison between data-driven percolation model and uniform model on a large sub-network with 41 million users of Twitter. The blue and orange symbols represent the simulation results with data-driven model M2 and uniform percolation model M5, respectively. The curves represent theoretical results corresponding to the simulation results with the same color. (**D**) Relation between percolation threshold and activity exponent $\alpha$ for different networks. The pink and blue curves are the theoretical results for varying $\alpha$ in Weibo and Twitter respectively. The green and yellow curves represent the results for directed ER networks and directed SF-ER networks respectively (see Supplementary Figure 29). The interval between the two blue square dots contains the empirical results of $\alpha$ in Twitter. See Supplementary Method 5 for the calculation of empirical critical points. (**E**) The percentage of top users who possess the most of influence power in Twitter. The curve represents the simulation results of model M2 with cascade size $P_\infty = 30000$. The imbalance tends to become more prominent when $\alpha$ increases, which is consistent with the result in Weibo. All error bars represent standard deviations. We use log scale for both axis in A and B, set only x-axis as log scale in C, and set only y-axis as log scale in D.

**Table 1: The collected network datasets of different social media.** In the fourth column, the symbol ✓ indicates that we have collected the friendship network structure between users, while the symbol ✗ indicates that we have only collected the degrees $k_i$ and $k_o$ of users.

| Network | Collected time | Network size | Network structure |
| --- | --- | --- | --- |
| Weibo 2014 | Mar. to Aug. 2014 | 99,546,027 | ✓ |
| Weibo 2013 | Nov. 4th, 2013 | 185,447 | ✗ |
| Weibo 2012 | Dec. 21st, 2012 | 309,091 | ✗ |
| Twitter 2009 [43] | Jul. 2009 | 41,652,230 | ✓ |

**Table 2: The collected evolution datasets of user in different social media.** The collected profile data contains the user identity, the follower count $k_o$, the followee count $k_i$ and the tweet count $s$. The users of evolution dataset 'Weibo 2013' is a part of the users of evolution dataset 'Weibo 2014'. The standard deviations of the exponents and the 95% confidence intervals (CI) are obtained by the nonparametric bootstrap method [56].

| Dataset | Size | Staring time | Ending time | $\alpha$ | $\sigma$ | $\tau$ |
|---|---|---|---|---|---|---|
| Weibo 2014 | 192,749 | Nov. 4th, 2013 | May 21st, 2014 | $0.284 \pm 0.017$ CI: $[0.258, 0.316]$ $p < 0.001$ $-0.60 \pm 0.19$ CI: $[-1.28, -0.14]$ $p = 0.05$ | $1.11 \pm 0.04$ CI: $[1.0, 1.18]$ $p < 0.001$ | $0.27 \pm 0.07$ $[0.10, 0.38]$ $p = 0.017$ |
| Weibo 2013 | 185,447 | Registration date | Nov. 4th, 2013 | $0.174 \pm 0.019$ CI: $[0.138, 0.216]$ $p < 0.001$ $-0.48 \pm 0.23$ CI: $[-1.16, -0.11]$ $p = 0.09$ | | |
| Weibo 2012 | 309,091 | Registration date | Dec. 21st, 2012 | $0.158 \pm 0.019$ CI: $[0.123, 0.198]$ $p < 0.001$ $-0.72 \pm 0.49$ CI: $[-1.28, 0.66]$ $p = 0.28$ | | |
| Twitter 2018 | 184,095 | Feb. 6th, 2018 | Mar. 3rd, 2018 | $0.68 \pm 0.05$ CI: $[0.57, 0.76]$ $p < 0.001$ $0.111 \pm 0.006$ CI: $[0.098, 0.122]$ $p < 0.001$ | $0.80 \pm 0.03$ CI: $[0.73, 0.87]$ $p < 0.001$ | $0.27 \pm 0.03$ $[0.20, 0.33]$ $p < 0.001$ |

**Table 3: The summary of percolation models.** The models M1 to M5 run on real social media networks, which are used in simulation. The models M6 to M10 are used in theoretical solution, in which only the real degree distribution $P(k_i, k_o)$ of social media network is used. The performance of M1, M2, M4, M6, M7, M9 are very close and no significant difference is found.

| Network | Nodal activity | Model notation |
|---|---|---|
| Real network structure | Real activity | M1 |
| | Fitting activity $m \sim m(k_o)$ | M2 |
| | Fitting activity $m \sim m(k_i)$ | M3 |
| | Fitting activity $m \sim m(k_i, k_o)$ | M4 |
| | Uniform activity $m =$ constant | M5 |
| Real degree distribution | Real activity | M6 |
| | Fitting activity $m \sim m(k_o)$ | M7 |
| | Fitting activity $m \sim m(k_i)$ | M8 |
| | Fitting activity $m \sim m(k_i, k_o)$ | M9 |
| | Uniform activity $m =$ constant | M10 |